\documentclass[conference,A4,twocolumn]{IEEEtran}
\usepackage[latin1]{inputenc}
\usepackage{psfrag}
\usepackage{graphicx}
\usepackage{amsmath}
\usepackage{amssymb}
\usepackage{amsbsy}
\usepackage{algorithmic}                       
\usepackage{algorithm}                         
\usepackage{subfigure}
\usepackage{color}
\begin{document}

\title{\huge Multi-Branch Lattice-Reduction SIC for
Multiuser MIMO Systems \vspace{-1em}}


\author{\IEEEauthorblockN{Leonel Ar\'evalo\IEEEauthorrefmark{1},
 Rodrigo C. de Lamare\IEEEauthorrefmark{1}, Keke Zu\IEEEauthorrefmark{2}
 and Raimundo Sampaio-Neto\IEEEauthorrefmark{1}}
 \IEEEauthorblockA{\IEEEauthorrefmark{1}{ Pontifical Catholic University of Rio de Janeiro (PUC-Rio)}\\
                                        { Center for Studies in Telecommunications (CETUC) --- Rio de Janeiro, Brazil}\\
                                        { E-mails: {\tt \{leonel\_arevalo, delamare and raimundo\}@cetuc.puc-rio.br}}}
 \IEEEauthorblockA{\IEEEauthorrefmark{2}{ Ericsson Research, Sweden}\\
                                        { E-mail: {\tt zukeke@gmail.com}}}
}

\maketitle

\begin{abstract}
In this paper, we propose a new detection technique for multiuser
multiple-input multiple-output (MU-MIMO) systems. The proposed
scheme combines a lattice reduction (LR) transformation, which makes
the channel matrix nearly orthogonal, and then employs a
multi-branch (MB) technique with successive interference
cancellation (SIC). A single LR transformation is required for the receive filters of all branches in the scheme, which proposes a different
ordering for each branch and generates a list of detection
candidates. The best vector of estimated symbols is chosen according
to the maximum likelihood (ML) selection criterion. Simulation
results show that the proposed detection structure has a
near-optimal performance while the computational complexity is much
lower than that of the ML detector.
\end{abstract}


\section{Introduction}
\label{intro}
Multi-input multi-output (MIMO) systems are key to increasing the
spectral efficiency of wireless communication systems, which makes
high data rates transmission possible~\cite{ref5}. Despite its
advantages, MIMO suffers from interference between multiple antennas
and, in multiuser scenarios, is affected by the multiuser
interference. To ensure the viability of MIMO systems it is
necessary to develop efficient detection techniques with low
computational complexity. In this direction many detectors have been
proposed in the last decade or so, see,
e.g.,~\cite{ref8}\nocite{ref6}\nocite{ref9}\nocite{ref10}--\cite{ref7}.
In ~\cite{ref1}  the complex lattice reduction (LR) was proposed to
improve the performance of MIMO detection techniques. The complex LR
transformation finds a new basis for the channel matrix, which is
nearly orthogonal allowing a more effective detection. The ordered
successive interference cancellation (OSIC)~\cite{ref2} can offer a
good performance when combined with other detection schemes. The
lattice reduction successive interference cancellation (LR-SIC) has
been developed in~\cite{ref4},\cite{dfjio},\cite{mfsic},\cite{mfdf},
\cite{did}. The works in \cite{ref14} and \cite{ref3} proposed a
novel successive interference cancellation (SIC) strategy for
communications systems based on multiple branches (MB). Although
LR-based detectors can attain the full receive diversity, a key
problem with these detectors \cite{ref8}--\cite{ref7} is that the
performance gap to the ML detector increases with the system size
and the modulation order.

In this work we propose a new detection scheme for multiuser MIMO
(MU-MIMO) systems in the uplink scenario that combines LR, SIC and
MB schemes to devise an MB-LR-SIC detector and which obtains a
near-ML performance. The main idea is to employ an LR transformation
in the channel matrix, then generate multiples branches, where each
branch has a different ordering pattern and produces different
symbol estimate vectors using the LR-SIC structure. In particular,
the first branch employs a column-norm based ordering in the LR
domain and the remaining branches use shifted versions of the order
of the first branch. The branch with the best performance among the
list of candidates is selected using the ML criterion. We study the
performance of the proposed and existing algorithms using a
realistic MU-MIMO scenario with path loss, log-normal shadowing and
multiple antennas per user. We also assess the performance of the
proposed MB-LR-SIC detector with practical channel estimation
algorithms and compare with with several existing detectors.
Simulation results show that the system performance with the
proposed MB-LR-SIC detector approaches the optimal ML detector.

This paper is structured as follows. Section~\ref{sec:Sect_I}
examines the basic concepts for multiuser MIMO system. The proposed
MB-LR-SIC detection scheme is detailed in Section~\ref{sec:Sect_II}.
Section~\ref{sec:results} is dedicated to the presentation and
discussion of the numerical results obtained via computer
simulations. Section~\ref{sec:conclusions} summarizes the
conclusions.

\section{System Model and Existing Techniques}
\label{sec:Sect_I} In this section, the basic concepts for multiuser
MIMO systems are studied, starting with the mathematical
representation of MU-MIMO systems and then a channel estimation
technique for MIMO channels is described. We also review two
important detection techniques, namely, complex lattice reduction
SIC and multi-branch SIC.


\subsection{System Model}

We consider the uplink of a multiuser MIMO system, as depicted in
Fig.~1, with $K$ active users, $N_{t_k}$ antennas at the $k$-th
mobile station (MS), and $N_r$ antennas at the base station (BS),
where $ N_r\geqslant \sum_{k=1}^KN_{t_k}$. The received signal
vector at BS is given by
\begin{eqnarray}
\mathbf{y}&=&  \mathbf{H}_1\mathbf{s}_1+
\mathbf{H}_2\mathbf{s}_2+ \ldots +
\mathbf{H}_K\mathbf{s}_K+\mathbf{n},
 \label{eq1}
\end{eqnarray}
where $\mathbf{s}_k$ is the $N_{t_k}\times 1$ transmitted signal
vector by the $k$-th user taken from a modulation constellation
$\mathcal{A}=\{a_1, a_2, \ldots, a_M \}$, each symbol is carrying
$C$ bits and $M=2^C$. $\mathbf{H}_k$ is the $N_r \times N_{t_k} $
channel matrix of the $k$-th user with elements $h_{n_r,n_{t_k}}$
correspond to the complex channel gains from the $n_{t_k}$-th
transmit antenna to the $n_r$-th receive antenna. The vector
$\mathbf{n}$ is an $N_r \times 1$ zero mean complex circular
symmetric Gaussian noise vector with covariance matrix
$\mathbf{K}_\mathbf{n}=\mathbb{E}[\mathbf{n}\mathbf{n}^{\mathcal{H}}]=\sigma_n^2\mathbf{I}$,
where $\mathbb{E}[\cdot]$ and $(\cdot)^\mathcal{H}$ represents the
expected value and Hermitian operator respectively. The expression
in (\ref{eq1}) can be written more conveniently as
\begin{eqnarray}
\mathbf{y}&=&\mathbf{H}\mathbf{s}+\mathbf{n}. \label{eq2}
\end{eqnarray}
where $\mathbf{H}=[\mathbf{H}_1 \,\,\, \mathbf{H}_2 \ldots
\mathbf{H}_K ]$ and $\mathbf{s}=[\mathbf{s}_1^T \,\,\,
\mathbf{s}_2^T \ldots \mathbf{s}_K^T ]^T$, $(\cdot)^T$ denotes
transpose operator. The symbol vector $\mathbf{s}$ of all $K$ users
has zero mean and a covariance matrix
$\mathbf{K}_\mathbf{s}=\mathbb{E}[\mathbf{s}\mathbf{s}^\mathcal{H}]=\sigma_s^2\mathbf{I}$,
where $\sigma_s^2$ is the signal power.

\begin{figure}
\begin{center}
\includegraphics[scale=0.28]{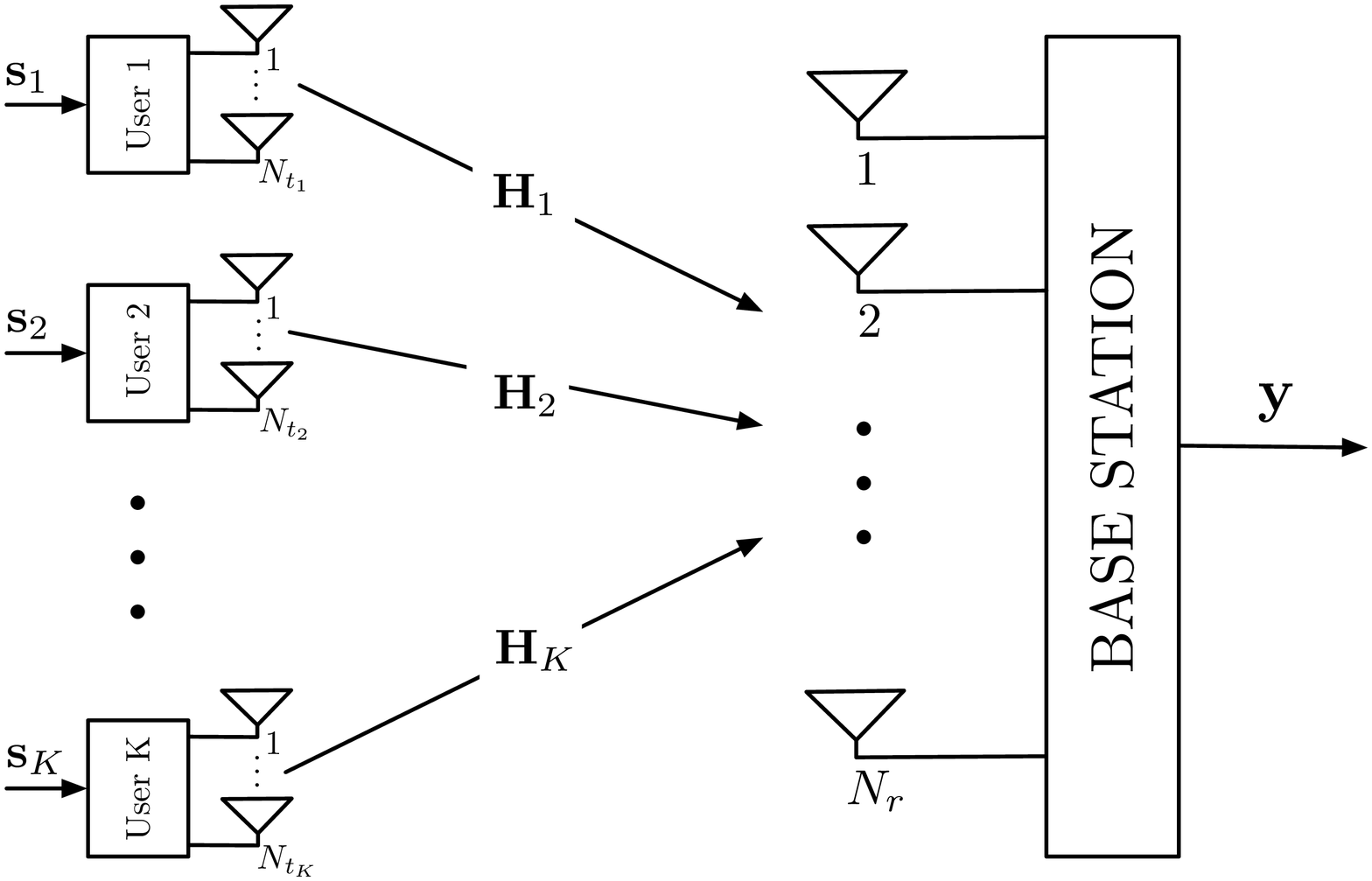}
\caption{Multiuser MIMO system.}
\end{center}
\label{MiMo_Diag}
\end{figure}

\subsection{MIMO Channel Estimation}

In this paper the LS channel estimation algorithm for MIMO system
proposed in~\cite{ref12} is used. In LS algorithm the cost function
in the time instant $i$ must be defined based on a weighted average
of error squares as
\begin{equation}
\mathcal{J}[i]=\sum_{k=1}^{i}\lambda^{i-k}\parallel \mathbf{y}[k]-\hat{\mathbf{H}}\mathbf{s}[k]\parallel^2,
\end{equation}
where $\mathbf{y}[k]$ and $\mathbf{s}[k]$ are the received and
transmitted symbol vectors in the time instant $k$, respectively,
$\lambda$ is the forgetting factor, $\hat{\mathbf{H}}[i]$ is the
channel matrix estimate in the time instant~$i$. The cost function
is minimized by solving
$\nabla_{\hat{\mathbf{H}}[i]}\mathcal{J}[i]=\mathbf{0}_{N_r,N_t}$,
where $\mathbf{0}_{N_r,N_t}$ denotes the $N_r \times N_t$ zero
matrix. The LS estimation of channel matrix can be solved easily as
\begin{equation}
\hat{\mathbf{H}}[i]=\mathbf{D}[i]\mathbf{P}[i],
\end{equation}
where $\mathbf{D}[i]$ can be iteratively calculated by
\begin{equation}
\mathbf{D}[i]=\lambda \mathbf{D}[i-1]+\mathbf{y}[i]\mathbf{s}[i]^\mathcal{H}.
\end{equation}
The matrix $\mathbf{P}[i]$ can be calculated iteratively by using
the matrix inversion lemma as
\begin{equation}
\mathbf{P}[i]=\lambda^{-1}\mathbf{P}[i-1]-\frac{\lambda^{-2}\mathbf{P}[i-1]\mathbf{s}[i]\mathbf{s}[i]^\mathcal{H}\mathbf{P}[i-1]}{1+\lambda^{-1}\mathbf{s}[i]^\mathcal{H}\mathbf{P}[i-1]\mathbf{s}[i]}.
\end{equation}
Initially, we set the parameters
$\mathbf{D}[0]=\mathbf{0}_{N_r,N_t}$ and
$\mathbf{P}[0]=\delta^{-1}\mathbf{I}$ where $\delta$ is a small
constant \cite{jio},\cite{jidf},\cite{l1stap}.

\subsection{Complex Lattice Reduction SIC Algorithm}
\label{L-R-SIC}

The complex lattice reduction scheme finds a new basis
$\mathbf{\tilde{H}}=\mathbf{HT}$ which is shorter and nearly
orthogonal compared to the original matrix $\mathbf{H}$, where
$\mathbf{T}$ is a uni-modular matrix ($\text{det}\mid
\mathbf{T}\mid$=1) and all elements of $\mathbf{T}$ are Gaussian
integers, i.e. $t_{i,i}\in \mathcal{Z}+j\mathcal{Z}$. Linear
detectors such as Zero Forcing (ZF) or minimum mean square error
(MMSE) can be developed on the LR transformed channel matrix
$\tilde{\mathbf{H}}$. Then satisfactory results have been obtained
due to reduced noise enhancement obtained with $\tilde{\mathbf{H}}$
\cite{ref4},\cite{lcbd},\cite{glcbd}.

With the channel matrix in the LR domain,
$\mathbf{\tilde{H}}=\mathbf{HT}$ and the definition of
$\mathbf{z}=\mathbf{T^{-1}s}$, the received signal vector
in~(\ref{eq2}) can be rewritten as
\begin{equation}
\mathbf{y}=\mathbf{H}\mathbf{s}+\mathbf{n}=\mathbf{HTT^{-1}s}+\mathbf{n}=\mathbf{\tilde{H}}\mathbf{z}+\mathbf{n}.\label{eq3}
\end{equation}
Even though $\mathbf{\tilde{H}}$ is nearly orthogonal, mutual
interference between the components of the transformed signal
$\mathbf{z}$ is still present. For this reason SIC techniques in the
LR domain results in additional improvements. The LR-SIC receiver
consists of a bank of linear detectors based on the matrix
$\mathbf{\tilde{H}}$, each detects a selected component of
$\mathbf{z}$. The component obtained by the first detector is used
to reconstruct the corresponding signal vector which is then
subtracted from the received signal to further reduce the
interference in the input to the next linear receive filter:
\begin{equation}
\mathbf{y}_1=\mathbf{y}-\mathbf{\tilde{h}}_i\hat{z}_i.
\end{equation}
The estimate $\hat{z}_i$ of  ${z}_i$ is obtained after shifting and
scaling operations~\cite{ref13} in the output of the $i$-th linear detector $\tilde{z}_i$, it is necessary to introduce re-scaling and re-shifting operations over the symbols, obtained by
\begin{equation}
\hat{z}_i=2\lfloor \frac{\tilde{z}_i-\kappa\mathbf{t}^{-1}_i\mathbf{1}}{2}  \rceil + \kappa \mathbf{t}_i^{-1}\mathbf{1}, \,\,\, i=1,\ldots, N_t,
\end{equation}
where $\mathbf{1}$ is a vector of ones, $\kappa =1+j$ for all M-QAM modulation, $\lfloor\cdot \rceil$ denotes the rounding operator and $\mathbf{t}^{-1}_i$ is the $i$-th row of $\mathbf{T}^{-1}$.
The index of $\hat{z}_i$ depends on the selected order in which the
interference vectors are subtracted. Then a new
$\mathbf{\tilde{H}}_1$ is calculated
\begin{equation}
\mathbf{\tilde{H}}_1=[\mathbf{\tilde{h}}_1\,\,\,\,\mathbf{\tilde{h}}_2 \ldots \mathbf{\tilde{h}}_{i-1}\,\,\,\, \mathbf{\tilde{h}}_{i+1} \ldots \mathbf{\tilde{h}}_{N_t} ],
\end{equation}
where $N_t=\sum_{k=1}^K N_{t_k}$. The second linear detector uses
the matrix $\mathbf{\tilde{H}}_1$, and the process is repeated $N_t$
times until all components of the estimated vector
$\mathbf{\hat{z}}$ are found. Finally, the estimated transmit signal
vector in the constellation domain can then be obtained by
\begin{equation}
\mathbf{\hat{s}}=\mathbf{T}\mathbf{\hat{z}}.\label{eq5}
\end{equation}
The main steps to find $\tilde{\mathbf{H}}$ and $\mathbf{T}$ with
CLLL reduction algorithm are described in~\cite{ref1}.

\subsection{Multi-Branch SIC Detection}

In the multi-branch scheme different orderings are explored for SIC,
each ordering is referred to as a branch, so that a detector with
$L$ branches produces a set of $L$ estimated vectors. Each branch
uses a column permutation matrix $\mathbf{P}$.  The estimate of the
signal vector of branch $l$, $\mathbf{\hat{x}}_l$, is obtained using
a SIC receiver based on a new channel matrix $\mathbf{H}^{(l)}
=\mathbf{H}\mathbf{P}_l$. The order of the estimated symbols is
rearranged to the original order by
\begin{equation}
\mathbf{\hat{s}}_l=\mathbf{P}_l\mathbf{\hat{x}}_l,  \,\,\,\,\,\,\, l=1, \ldots, L.\label{eq4}
\end{equation}
A higher detection diversity can be obtained by selecting the most
likely symbol vector based on the ML selection rule.

\begin{figure}
\begin{center}
\includegraphics[scale=0.31]{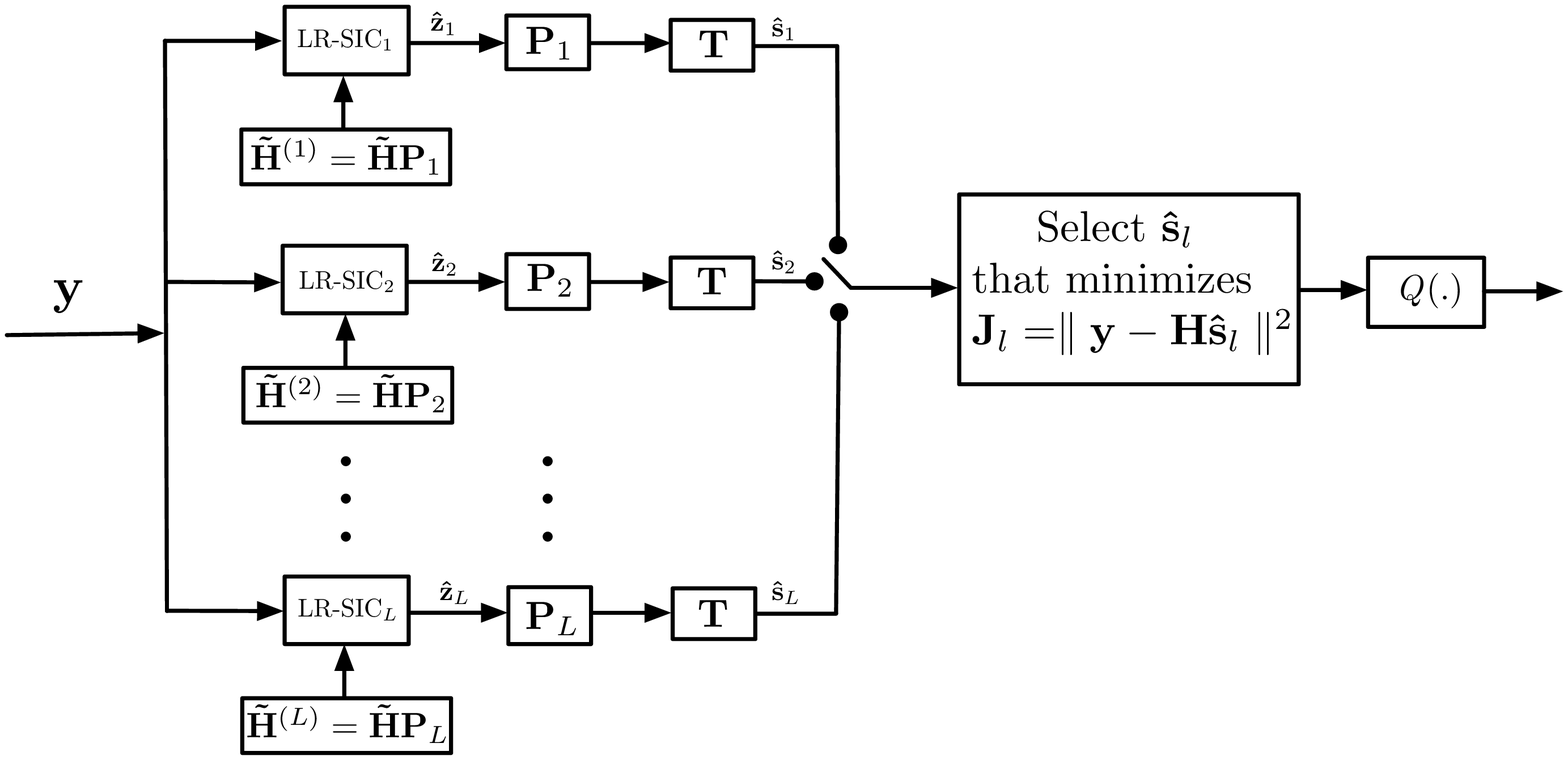}
\label{LR-MB-SIC-Diag} \caption{Proposed multi-branch lattice
reduction successive interference cancellation detection block
diagram.}
\end{center}
\end{figure}

\section{Proposed Multi-Branch Lattice Reduction SIC Detection}
\label{sec:Sect_II}
This section describes the proposed MB-LR-SIC detection scheme that
combines the concepts previously presented. A schematic of the
detector is shown in Fig. 2. The proposed detection structure
employs an LR transformation in the channel matrix as
in~(\ref{eq3}). Let $\mathbf{P}_l$ be a column permutation matrix.
Then the signal term in~(\ref{eq3}) can be expressed as
\begin{equation}
\mathbf{\tilde{H}z}=\mathbf{\tilde{H}P}_l\mathbf{P}_l^T\mathbf{z}=\mathbf{\tilde{H}}^{(l)}\mathbf{z}_l,
\end{equation}
where
$\mathbf{\tilde{H}}^{(l)}=\mathbf{\tilde{H}P}_l=\mathbf{H}(\mathbf{T}\mathbf{P}_l)$
is a column permutated version of $\mathbf{\tilde{H}}$ and
\begin{equation}
\mathbf{z}_l=\mathbf{P}_l^T\mathbf{z}=(\mathbf{P}_l^T\mathbf{T}^{-1})\mathbf{s}.
\end{equation}
In the proposed scheme, the $l$-th branch ($l=1,\ldots,L$) of the
receiver employs an LR-SIC detector, as described in
Section~\ref{L-R-SIC}, based on the permutated matrix
$\mathbf{\tilde{H}}^{(l)}=\mathbf{\tilde{H}}\mathbf{P}_l$ to
generate an estimate $\mathbf{\hat{z}}_l$ of $\mathbf{z}_l$. Then, the $l$th branch
transmitted signal candidate is obtained by
\begin{equation}
\mathbf{\hat{s}}_l=(\mathbf{P}_l^T\mathbf{T}^{-1})^{-1}\mathbf{\hat{z}}_l=(\mathbf{T}\mathbf{P}_l)\mathbf{\hat{z}}_l, \hspace{0.5cm} l=1,\ldots,L.
\end{equation}
The best candidate out of the $L$ estimated data signal vectors is
selected using the ML criterion, that is
\begin{equation}
\mathbf{\hat{s}}= \text{arg min}\parallel \mathbf{y}-\mathbf{H}\mathbf{\hat{s}}_l \parallel^2, \hspace{0.5cm} l=1,\ldots,L.
\end{equation}
In order to guarantee that a good candidate is always present in the
set of the $L$ candidates generated, one of the branches (e.g.
$l=1$) should implement a performance effective ordering (e.g.
column-norm based ordering) in the LR domain. Here, the remaining
branches used the so called Pre-Stored Patterns (PSP) proposed
in~\cite{ref3}. The PSP can be described mathematically by
\begin{equation}
\mathbf{P}_l=\begin{bmatrix} \mathbf{I}_s & \mathbf{0}_{s,N_t-s} \\ \mathbf{0}_{N_t-s,s} & \phi[\mathbf{I}_s]\end{bmatrix},\,\,\, 2\leqslant l \leqslant N_t,
\end{equation}
where $\mathbf{0}_{m,n}$ denotes $m\times n$-dimensional matrix full of zeros, the operator $\phi[\cdot]$ rotates the elements of the argument matrix column-wise such that an identity matrix becomes a matrix with ones in the reverse diagonal. The PSP algorithm shifts the ordering of the cancellation according to shifts given by
\begin{equation}
s=\lfloor (l-2)N_t/L \rfloor, \,\, 2\leqslant l \leqslant N_t,
\end{equation}
where $L$ is the number of parallel branches and $\lfloor . \rfloor$ rounds the argument to the lowest integer according to the $l$-th branch.

The description of the proposed MB-LR-SIC detection
structure is depicted in Algorithm 1.

\begin{algorithm}{\small  
Initialization: $L=N_t$\\
$[\tilde{\mathbf{H}}, \mathbf{T}]=\text{CLLL}(\mathbf{H})$ \% \textit{CLLL in} \cite{ref1}, \textit{input} $\mathbf{H}$, \textit{output} $\tilde{\mathbf{H}}$ \textit{and} $\mathbf{T}$ \\
\textbf{Do for}{\,\,$l=1$ \textbf{to} $L$ \% \textit{Multi-Branch loop}}{\\
\hspace*{.4cm} \textbf{if}{\,\,\,$l=1$}\\
\hspace*{.8cm}$\mathbf{P}_l=\text{CNBO}(\tilde{\mathbf{H}})$ \% \textit{column-norm based ordering in} $\tilde{\mathbf{H}}$\\
\hspace*{.4cm} \textbf{else}\\
\hspace*{.8cm}$\mathbf{P}_l=\text{PSP}(N_t,l)$ \% \textit{Pre-Stored Patterns}\\
\hspace*{.4cm} \textbf{end if}\\
\hspace*{.4cm}$\tilde{\mathbf{H}}^{(l)}=\tilde{\mathbf{H}}\mathbf{P}_l$\\
\hspace*{.4cm}\textbf{Do for}{\,\,$n=1$ \textbf{to} $N_t$}\\
\hspace*{.8cm}$\mathbf{W}_{l,n}=\text{MMSE}(\tilde{\mathbf{H}}^{(l)})$ \% \textit{MMSE linear equalizer}\\
\hspace*{.8cm}$\tilde{\mathbf{z}}_{l,n}=\mathbf{W}_{l,n}^\mathcal{H}\mathbf{y}_{l,n}$\\
\hspace*{.8cm}$\hat{z}_{l,n}=\text{SS}(\tilde{z}_{l,1})$ \% \textit{Shifting and Scaling operations}~\cite{ref13}\\
\hspace*{.8cm}$\mathbf{y}_{l,n}=\mathbf{y}_{l,n}-\tilde{\mathbf{h}}^{(l)}_1\hat{z}_{l,n}$ \\
\hspace*{.8cm}$\tilde{\mathbf{H}}^{(l)}=[\tilde{\mathbf{h}}^{(l)}_2, \tilde{\mathbf{h}}^{(l)}_3, \ldots,\tilde{\mathbf{h}}^{(l)}_{N_t}]$\\
\hspace*{.4cm}\textbf{End}\\
\hspace*{.4cm}$\mathbf{s}_l=\mathbf{T}\mathbf{P}_l\hat{\mathbf{z}}_l$\\
\textbf{End}\\
$l_{opt}=\text{argmin}_{1\leq l\leq L}\parallel\mathbf{y}-\mathbf{H}\mathbf{s}_l\parallel^2$\\
$\hat{\mathbf{s}}=\mathbf{s}_{l_{opt}}$
}
\caption{\textbf{:}\hspace{.6cm} THE MB-LR-SIC ALGORITHM}
}

\end{algorithm}
The proposed scheme increases the
chances of obtaining more reliable candidates for the transmitted
data symbol vector. The MB-LR-SIC scheme shows a high diversity gain
and can deliver a performance very close to the optimal ML detector.

\begin{figure}[H]
\begin{center}
\includegraphics[scale=0.6]{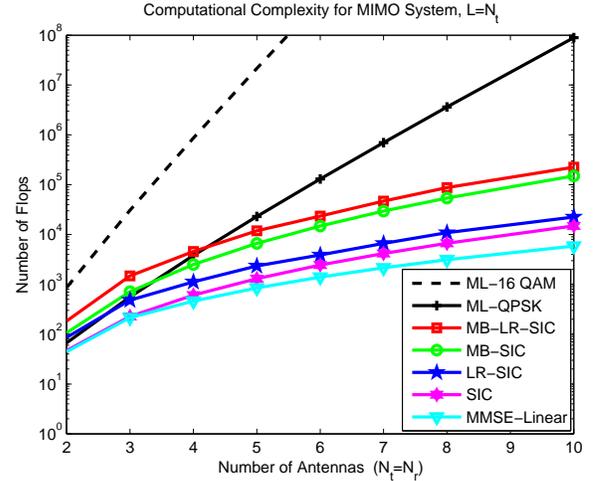}
\caption{Computational complexity of detection algorithms. }
\label{Complexity}
\end{center}
\end{figure}

\subsection{Computational complexity}

In this subsection, the computational complexity of the MB-LR-SIC is
evaluated. Since the LR  transformation requires the counting of
floating points (flops) to know its computational cost, we have
computed the number of flops per received vector $\mathbf{y}$ using
the Lightspeed Matlab toolbox~\cite{ref15}. In Fig.~\ref{Complexity}
it is shown the computational complexity for the different detection
algorithms studied in this work. The figure compares the required
number of flops versus the number of antennas for $N_t=N_r$. For the
ML detector, we have considered QPSK and 16-QAM modulation, whereas
for the other detectors we have used only QPSK modulation due to the
fact that the computational cost in these detectors does not vary
significantly with the modulation type. The proposed MB-LR-SIC has a
lower complexity than the ML detector and, as will be shown in the
next section, the performance is very close to the ML detector.

\section{Numerical Results}
\label{sec:results}

\begin{figure}
\begin{center}
\includegraphics[scale=0.6]{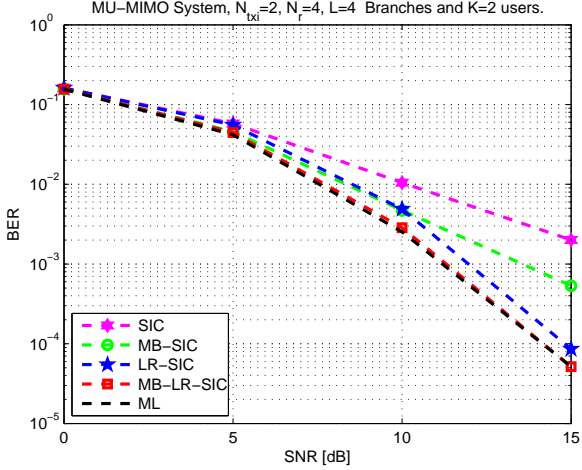}
\caption{BER vs SNR performance for the proposed  MB-LR- SIC and
existing algorithms in a MU-MIMO scenario. All SIC detectors use a
column-norm-based ordering and QPSK modulation.} \label{results1}
\end{center}
\end{figure}

\begin{figure}
\begin{center}
\includegraphics[scale=0.6]{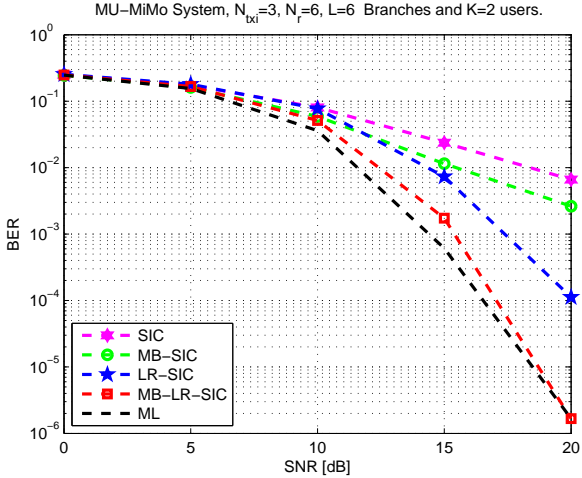}
\caption{BER vs SNR performance for the proposed  MB-LR- SIC and
existing algorithms in a MU-MIMO scenario. All SIC detectors use a
column-norm-based ordering and 16-QAM modulation.} \label{results2}
\end{center}
\end{figure}

In this section, the performance of the proposed MB-LR-SIC scheme is
compared with existing detection algorithms, which include the
standard SIC, LR-SIC, MB-SIC and ML. We consider the uplink of a
MU-MIMO system with $K$ active users. Two scenarios for the channels
associated with each active user are considered. In the first
scenario, we consider independent and identically distributed random
fading channel models whose coefficients are complex Gaussian random
variables with zero mean and unit variance. In the second scenario
we consider a more realistic channel described by
\begin{equation}
\mathbf{H}_k=\alpha_k\beta_k\mathbf{G}_k; \hspace{0.5cm} k=1,\ldots,K. \label{eq6}
\end{equation}
where $\alpha_k$ represents the distance based path-loss between the
$k$-th transmitter and the receiver, and $\beta_k$ is a log-normal
variable, representing the shadowing between the transmitter and the
receiver. The parameters $\alpha_k$ and $\beta_k$ are respectively
calculated by
\begin{equation}
\alpha_k=\sqrt{L_p^{(k)}},
\end{equation}
and
\begin{equation}
\beta_k=10^{\frac{\sigma_k \mathcal{N}_k(0,1)}{10}},
\end{equation}
where $L_p^{(k)}$ is the base power path loss, $\mathcal{N}_k(0,1)$
represents a Gaussian distribution with zero mean and unit variance
and $\sigma_k$ is the shadowing spread in dB. The matrix $\mathbf{G}_k$
in~(\ref{eq6}) is modeled as the Kronecker channel
model~\cite{ref11}, expressed by
\begin{equation}
\mathbf{G}_k=\mathbf{R}^{1/2}_{r_x}\mathbf{G}_{0_k} \mathbf{R}^{1/2}_{t_{x_k}},
\end{equation}
where $\mathbf{G}_{0_k}$ is the MIMO channel matrix for scenario~1
and $\mathbf{R}_{r_x}$ and $\mathbf{R}_{t_{x_k}}$ denote the receive
and transmit correlation matrices, respectively. Assuming
$L_p^{(k)}=L_p$, $\sigma_k=\sigma$ and the same correlation matrix
$\mathbf{R}^{1/2}_{t_{x_k}}=\mathbf{R}^{1/2}_{t_x}$ for all $K$
transmitters, the SNR is defined as $10\text{log}_{_{10}}\frac{N_{t}
\sigma_s^{2}}{\sigma_n^2}$, where $\sigma_s^2$ is the common
variance of the received symbols and $\sigma_n^2$ is the noise
variance.

\begin{figure}
\begin{center}
\includegraphics[scale=0.6]{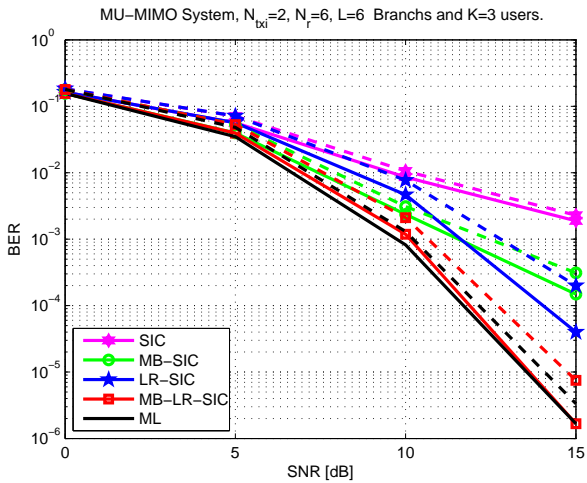}
\caption{Comparing BER vs SNR performance for the proposed  MB-LR- SIC and
existing algorithms in an MU-MIMO scenario with LS channel estimation (\textbf{- -}) and perfect channel estimation (\textbf{---}). }
\label{Estimation}
\end{center}
\end{figure}

\begin{figure}
\begin{center}
\includegraphics[scale=0.6]{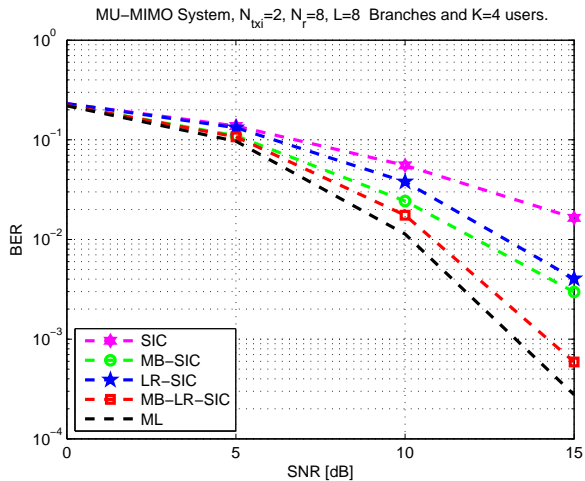}
\caption{BER vs SNR performance for the proposed MB-LR- SIC and
existing algorithms in an MU-MIMO scenario with realistic
propagation conditions. All SIC detectors use column-norm-based
ordering ($L_p=0.7, \sigma=6 \text{dB},
\rho_{t_x}=\rho_{r_x}=0.2$).} \label{results3}
\end{center}
\end{figure}

The simulation curves correspond to an average of 3,000 simulation
runs, with 500$N_t$ symbols transmitted per run. In
Fig.~\ref{results1} and Fig.~\ref{results2}, we consider scenario 1,
compare the performance of the proposed MB-LR-SIC detector with that
of existing detectors where SIC, MB-SIC and LR-SIC all use MMSE
linear receive filters, QPSK and 16-QAM modulation,
column-norm-based ordering and employ perfectly known channel state
information. Note that the performance of the MB-LR-SIC is very
close that of the optimum ML detector, even when the modulation
order and the number of antennas per user grow,
see~Figure.~\ref{results2}. In the next experiment, scenario 1 and
channel estimation for the different detectors is considered,
550$N_t$ symbols are transmitted, where 50 symbols are used for
training. In Fig.~\ref{Estimation} it is shown that the performance
of MB-LR-SIC with channel estimation (dotted lines) is still close
to the ML detector performance with channel estimation. The curves
in Fig.~\ref{results3} illustrate the behavior of the system
performance for the more realistic channel scenario with $L_p= 0.7$
and $\sigma=6$ dB. The components of correlation matrices
$\mathbf{R}_{r_x}$ and $\mathbf{R}_{t_{x_k}}$ are of the form:
\begin{equation}
\mathbf{R}=\begin{bmatrix}
1 & \rho & \rho^4 & \ldots & \rho^{(N_a-1)^2} \\ \rho & 1 & \rho & \ldots &
\vdots\\ \rho^4 & \rho & 1 & \vdots & \rho^4\\\vdots & \vdots & \vdots & \vdots & \vdots\\ \rho^{(N_a-1)^2} & \ldots & \rho^4 & \rho & 1
\end{bmatrix},
\end{equation}
where $N_a$ is the number of antennas and $\rho$ is the correlation
index of neighboring antennas ($\rho=\rho_{t_x}$ for the transmit
antennas and $\rho=\rho_{r_x}$ for the receive antennas). Note that
$\rho =0$ represents an uncorrelated scenario and $\rho =1$ implies
a fully correlated scenario. The curves in Fig.~\ref{results3}
display a loss of performance in all detectors due to the
propagation effects. However, the proposed MB-LR-SIC algorithm for
these realistic conditions outperforms the other considered
techniques and is close to the ML detector.

\section{Conclusions}
\label{sec:conclusions}
This paper introduced a new detection technique for multiuser MIMO
systems. The proposed MB-LR-SIC detector employs an LR
transformation technique and generates multiple ordering patterns
and estimates of symbol vectors, each of these ordering patterns
uses SIC detection in the LR domain. This makes possible to generate
a set of candidate solutions of the transmitted vector with high
reliability. Finally the best candidate is selected using the ML
criterion. The MB-LR-SIC detector, besides being conceptually
simple, has been shown through simulations that in a realistic
scenario and without perfect knowledge of the channel state
information (using channel estimation) the performance approaches
the optimal ML detector with much lower computational complexity.
Due to the higher detection diversity and low computational
complexity, compared with the optimal detector, the MB-LR-SIC
structure is presented as a viable detection alternative for future
MIMO communications systems.
\vspace{1cm}
\bibliographystyle{IEEEbib}
\bibliography{Referens}
\normalfont

\end{document}